\documentclass[journal]{IEEEtran}
\usepackage{graphicx}
\usepackage{epsf}
\usepackage{mathtools}
\usepackage{amssymb}
\usepackage{amsmath, amsfonts}
\usepackage{latexsym}
\usepackage{multirow}
\usepackage{multicol}
\usepackage[table]{xcolor}
\usepackage{color}

\usepackage{tabularx}
\usepackage{lipsum}

\usepackage{mathtools}

\usepackage{fixltx2e}

\usepackage[mathscr]{euscript}

\usepackage{commath}
\usepackage{MnSymbol}
\usepackage[]{footmisc}

\usepackage{mathtools} 
\DeclarePairedDelimiterX\setc[2]{[}{]}{\,#1 \;\delimsize\vert\; #2\,}
\DeclarePairedDelimiterX\parth[2]{(}{)}{\,#1 \;\delimsize\vert\; #2\,}

\PassOptionsToPackage{hyphens}{url}
\usepackage[unicode=true, bookmarks=true, bookmarksnumbered=true, bookmarksopen=true, bookmarksopenlevel=1, breaklinks=false, pdfborder={0 0 0}, pdfborderstyle={}, backref=false, colorlinks=false]{hyperref}

\usepackage[ruled,linesnumbered]{algorithm2e}

\usepackage{theorem}
{
\theorembodyfont{\rmfamily}

}

\usepackage{subcaption}
\usepackage{bbm}

\definecolor{orange}{RGB}{255,127,0}
\definecolor{blue}{RGB}{0,0,255}
\definecolor{red}{RGB}{255,0,0}
\definecolor{green}{RGB}{50,160,50}
\definecolor{grey}{RGB}{125,120,125}

\begin{document}

\title{BlueFMCW: Random Frequency Hopping Radar for Mitigation of Interference and Spoofing }

\author{Thomas Moon,~\IEEEmembership{Member,~IEEE,}
        Jounsup Park,~\IEEEmembership{Member, IEEE,}
        and~Seungmo Kim,~\IEEEmembership{Member, IEEE}}

\maketitle

\begin{abstract}
Traffic safety is the foremost value that automotive radar systems aim to pursue. Unlike in mobile communication systems, the literature for radar systems did not adequately address inter-radar interference and security threats such as jamming and spoofing, which in turn threatens the traffic safety. In this context, we introduce a novel frequency-modulated continuous-wave (FMCW) radar scheme (namely, BlueFMCW) that mitigates both interference and spoofing signals. BlueFMCW randomly hops frequency to avoid interference and spoofing signals. Our phase alignment algorithm is capable of removing the phase discontinuity while combining the beat signals from the randomly-hopped chirps, and thereby radar's resolution is not compromised. The simulation results show that BlueFMCW can efficiently mitigate the interference and spoofing signals in various scenarios without paying its resolution. 
\end{abstract}


\section{Introduction}\label{sec_intro}
Automotive sensor systems are considered as a key technology in autonomous driving by detecting surrounding obstacles. As a vital part of the sensing system, radars detect obstacles in a long-range, compared to cameras and ultrasonic sensors focused on shorter-range detection. In particular, radar techniques \cite{radaridentity} using radio frequency (RF) have garnered attraction mainly thanks to its capability of detecting objects in various climate conditions. Among those, frequency-modulated continuous-wave (FMCW) allows the radar to detect both range and speed of objects \cite{FMCWLidar, FMCWLidarInterference}, which raises considerable research interest.

\begin{figure}[t]
\begin{center}
\includegraphics[width=0.4\textwidth]{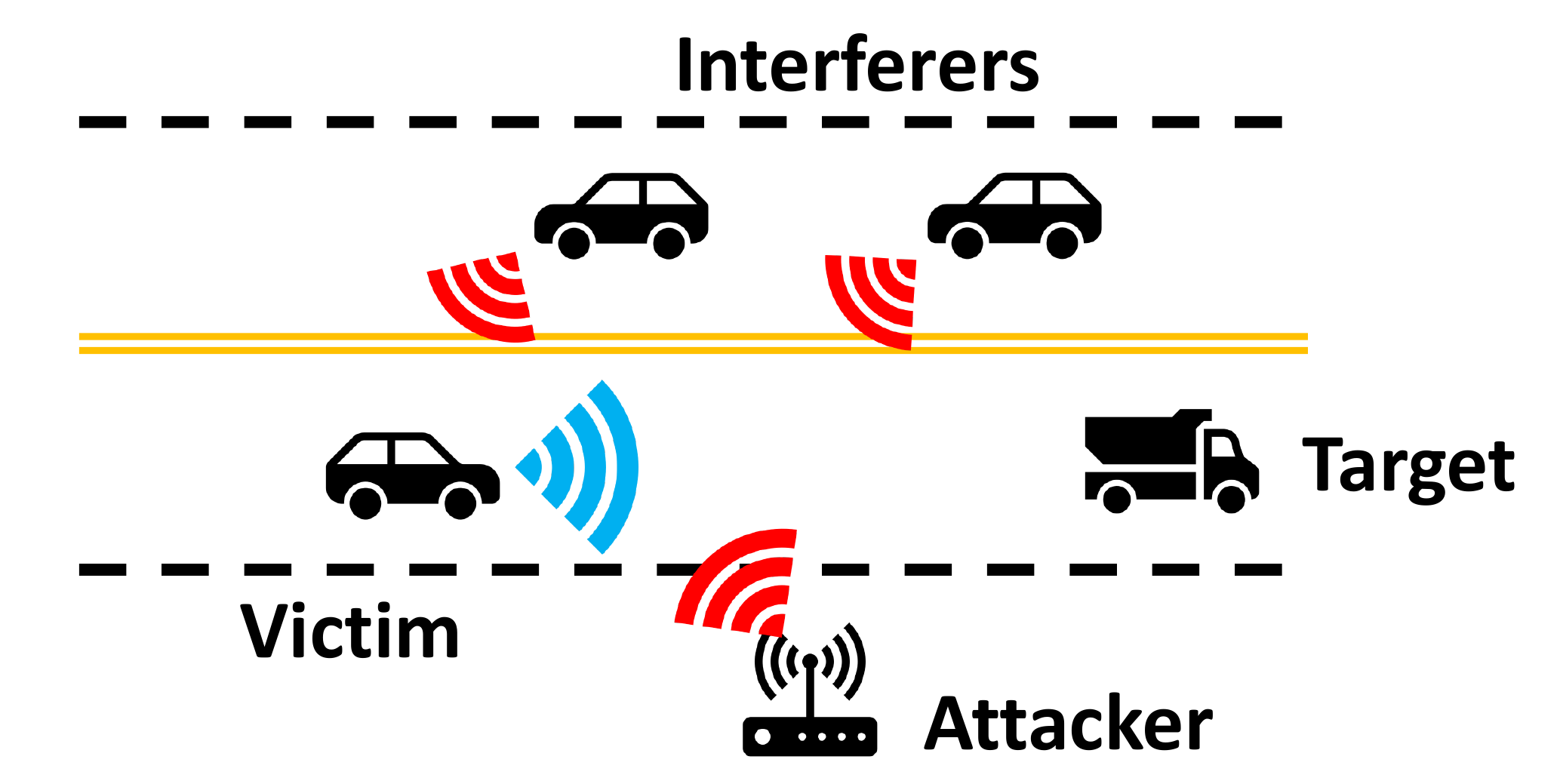}\\
\caption{ \textbf{Threat models in automotive radar system.} } 
\label{fig:intro}
\end{center}
\vskip -0.25in
\end{figure}

However, as more vehicles are equipped with radars, the interference will likely become a salient issue to resolve (as illustrated in Figure.~\ref{fig:intro}). An increasing number of interfering signals between radars increases the noise floor, degrading the radars' accuracy. Thus, mitigation of the interference is increasingly critical in keeping the performance of a radar operable \cite{alland2019interference}. Radar interference can threaten the safety of autonomous vehicles in the near future, as 50\% of new vehicle sales will be autonomous in the 2050s \cite{litman2017autonomous}. Autonomous vehicles require radars for adaptive cruise control (ACC), pedestrian detection, cross-traffic alerts, blind spot detection (BSD), rear collision warning, etc. These features are closely related to the safety of the vehicles.

Another significant aspect to consider in the design of an automotive radar is \textit{security}. There are two principle attacks (i.e., intentional disruption of a vehicular system by a third-party) on an automotive radar. First, \textit{jamming} attacks have the potential for inducing major collisions in the future. Jamming attackers intentionally send out high power radio signal to confuse or overwhelm the radar receiver. Jamming attacks make the radar out of operation by saturating the radar with noise. However, the damage by a jamming attack can be reduced in a highly mobile environment, making it very difficult for an attacker to target specific vehicles \cite{heath}. Second, automotive radars are known to be susceptible to \textit{spoofing}, replicating and retransmitting radar transmit signals designed to provide false information to radar to corrupt received data. Spoofing on distance and velocity on commercial automotive radars have been shown to be feasible \cite{chaun14}.

As the attacker is likely to use similar hardware with the victim automotive radar, it can listen for the original radar signal and generate a similar radar signal to confuse the victim radar. If the attacker generates a noise-like jamming signal to overwhelm the signal, the victim radar can detect the attack and switch to other types of sensors to continue driving. However, if the attacker does spoofing by transmitting a similar signal that is not distinguishable from the original radar signal, it can be a threat to vehicles' safety on the road. The \textit{ghost} objects generated by the spoofing can cause the vehicle to stop to avoid a collision. It can cause accidents or a non-necessary slow down of traffic.

\subsubsection{Related Work}
There are several prior works on mitigation techniques against mutual interference or attacks on the automotive radars. \textit{Communication-based} techniques such as dual function radar-communication \cite{ma2019joint}\cite{huang2020majorcom} proved effectiveness only against interference signals. Meanwhile, our proposed work provides a countermeasure against intended attack signals as well.

As efforts to combat the radar interference or attacks, techniques such as \textit{beamforming} \cite{bechter2016digital}\cite{bechter2017analytical} and \textit{polarization} \cite{dai2012main}\cite{dai2011novel} are also found in the literature.

Further enhancements on the FMCW are found in the latest work such as \textit{adaptive noise cancellation} techniques \cite{gerstmair2019safe, jin2019automotive}, which improves the signal-to-noise-ratio by canceling the noise from the radar interference. However, they did not specify how to distinguish false objects created by the interference or attack, which is the main problem that our work focuses to address.

\textit{Phase-Coded FMCW (PC-FMCW)} \cite{uysal2019pcfmcw} is proposed to mitigate interference and enable joint sensing and communication. While the phase coding enables carrying information and mitigating the interference simultaneously, it causes phase discontinuity on the waveforms. The realization of such waveforms in hardware is challenging due to the instantaneous phase change needing costly equipment; thus, it does not apply to the automotive radar domain straightforward. 

To address the limitations of the above-mentioned techniques, we introduce ``BlueFMCW,'' a novel FMCW mechanism that features robustness against interference and spoofing attacks simultaneously. The technique is named after Bluetooth \cite{bluetooth} for the frequency hopping to avoid interfering signals. Bluetooth randomly switches the frequency over 79 channels between 2400 and 2483.5 MHz at a rate of 1600 hops per second. We adopt the frequency hopping to a radar system, in which way multiple radars can share a same band without interference thanks to the random hopping pattern. It also makes the BlueFMCW robust against spoofing attacks by making attackers not easily able to decrypt key parameters such as a random hopping pattern.

The idea of using a random waveform to mitigate the interference has been proposed before this paper. A recent work proposed an amplitude-modulated chirp waveform for a transmitted signal \cite{guan2019application}. It exploits a hash function for the amplitude modulation and the pulse repetition period to avoid the interfering signal. Nonetheless, due to reliance on a precise control of the amplitude, its performance under noise cannot be guaranteed. Another recent proposal introduced a FMCW mechanism based on a random chirp signal \cite{liu2018high}. It divides the pulse into $N$ smaller chirp chips while randomly switching the starting frequencies of each chip. The limitation is that the resolution is sacrificed by the number of sub-chirps, focusing on confusing attackers by hopping the frequency and using the information from a single chirp chip. It still leaves the proposal vulnerable to a spoofing attack.

To this end, we use a frequency hopping method where all the best signals from the sub-chirps are aggregated. It enables us to achieve $N$ times higher resolution than using a single chirp chip. Our work also effectively suppresses the in-band spoofing or interference signal by spreading out their energy.

\subsubsection{Contributions}
We summarize the contributions of BlueFMCW as follows. 

\begin{itemize}
    \item BlueFMCW efficiently avoids both interference signals and spoofing attacks.
    \item BlueFMCW does not degrade the resolution because of the random frequency hopping.
    \item Phase alignment algorithm is provided to resolve the phase discontinuity problem caused by random frequency hopping.
    \item BlueFMCW is compatible with conventional FMCW radars.
\end{itemize}

\subsubsection{Organization}
This paper is organized as follows. Section II covers the background of FMCW radar and adversary model. Section III details the design of BlueFMCW radar. Sections IV presents performance evaluation of BlueFMCW radar compared with baseline methods. Section V. related works, and Section VI concludes the paper.
\section{Background}\label{sec_background}
\begin{figure}[t]
\begin{center}
\includegraphics[width=0.4\textwidth]{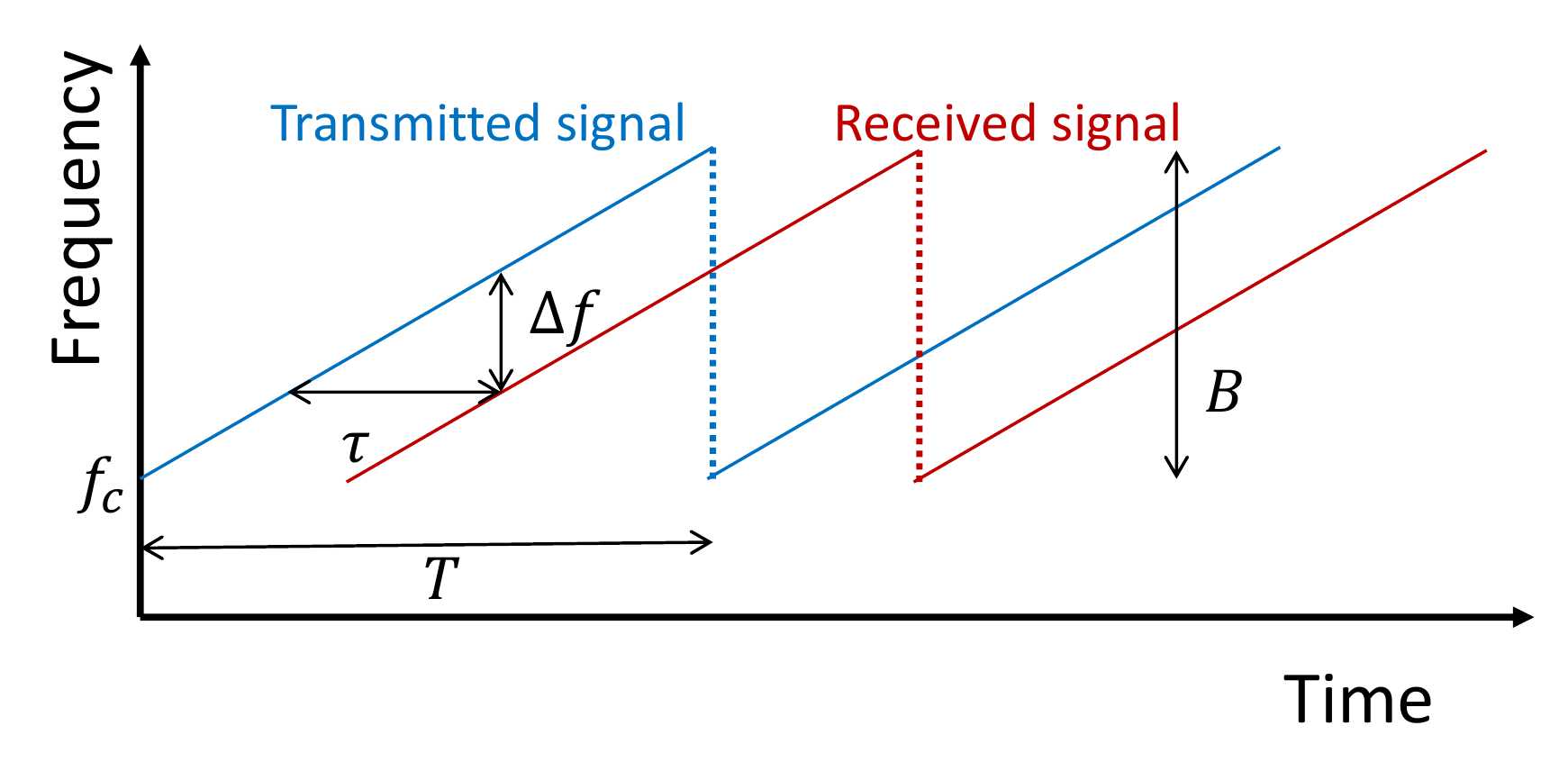}\\
\caption{ \textbf{Ranging distances using FMCW.} The figure shows the transmitted FMCW signal (blue) and its reflection (red). The time-of-flight ($\tau$) between the two signals maps to the frequency shift $\Delta f$.  } 
\label{fig:fmcw}
\end{center}
\vskip -0.25in
\end{figure}

\subsection{FMCW radar}
FMCW radar continuously transmits periodic pulses whose frequency sweeps linearly in time, as shown by the blue line in Figure~\ref{fig:fmcw}. Mathematically, the transmitted signal is 
\begin{align}
s_t(t) = \exp(j2\pi (f_c t + \frac{\alpha}{2}t^2))
\end{align}
where $f_c$ and $\alpha$ are the starting frequency and slope of the FMCW chirp, respectively. The reflected signal is a time-delayed version of the transmitted signal, which arrives after bouncing off a reflector, as illustrated by the red line shown in Figure~\ref{fig:fmcw}. The time-of-flight (TOF, $\tau$) is an elapsed time for the transmitted signal, traveling the round-trip distance $2d$ from the radar to the reflector and back to the radar. In the presence of multiple reflectors, the received signal is written as $s_r(t) = \Sigma A_i s_t(t-\tau_i)$, where $A_i$ and $\tau_i$ are the attenuation factor and TOF of the $i$th reflector. The receiver mixes the transmitted signal with the received signal to produce a beat signal $x(t) = s_r(t)\cdot s_t(t)^*$. Plugging the above two equations, the beat signal becomes
\begin{align} \label{eq:beat}
x(t) = \sum_i A_i \exp(j2\pi (\alpha \tau_i t + f_c \tau - \frac{\alpha}{2}\tau_i^2)).
\end{align}
This enables us to extract a distance profile of multiple reflectors because the frequency of the beat signal is $\alpha \tau$ where $\alpha$ is a known parameter. Hence, we can perform the FFT on the beat signal and detect the objects by finding the peaks. 

The resolution of FMCW is defined by the ability to distinguish two objects close to each other. Since their distances are proportional to their frequency shifts $\Delta f$, the resolution of distance $\Delta d$ is a function of the ability to distinguish two frequency shifts. The resolution of differentiating two frequencies $\Delta f_{min}$ is equal to the size of a single FFT bin. When the FFT is taken over a chirp-duration $T$, the size of one FFT bin is equal to $1/T$. Hence, the distance resolution is $\Delta d = \frac{C \cdot \Delta f_{min}}{2\alpha} = \frac{C}{2\alpha T} = \frac{C}{2 B}$, where $B$ is the frequency bandwidth equal to $\alpha T$. It is important to note that the resolution of FMCW is a function of the bandwidth of the transmitted signal. 

\begin{figure}[t]
\begin{center}
\includegraphics[width=0.45\textwidth]{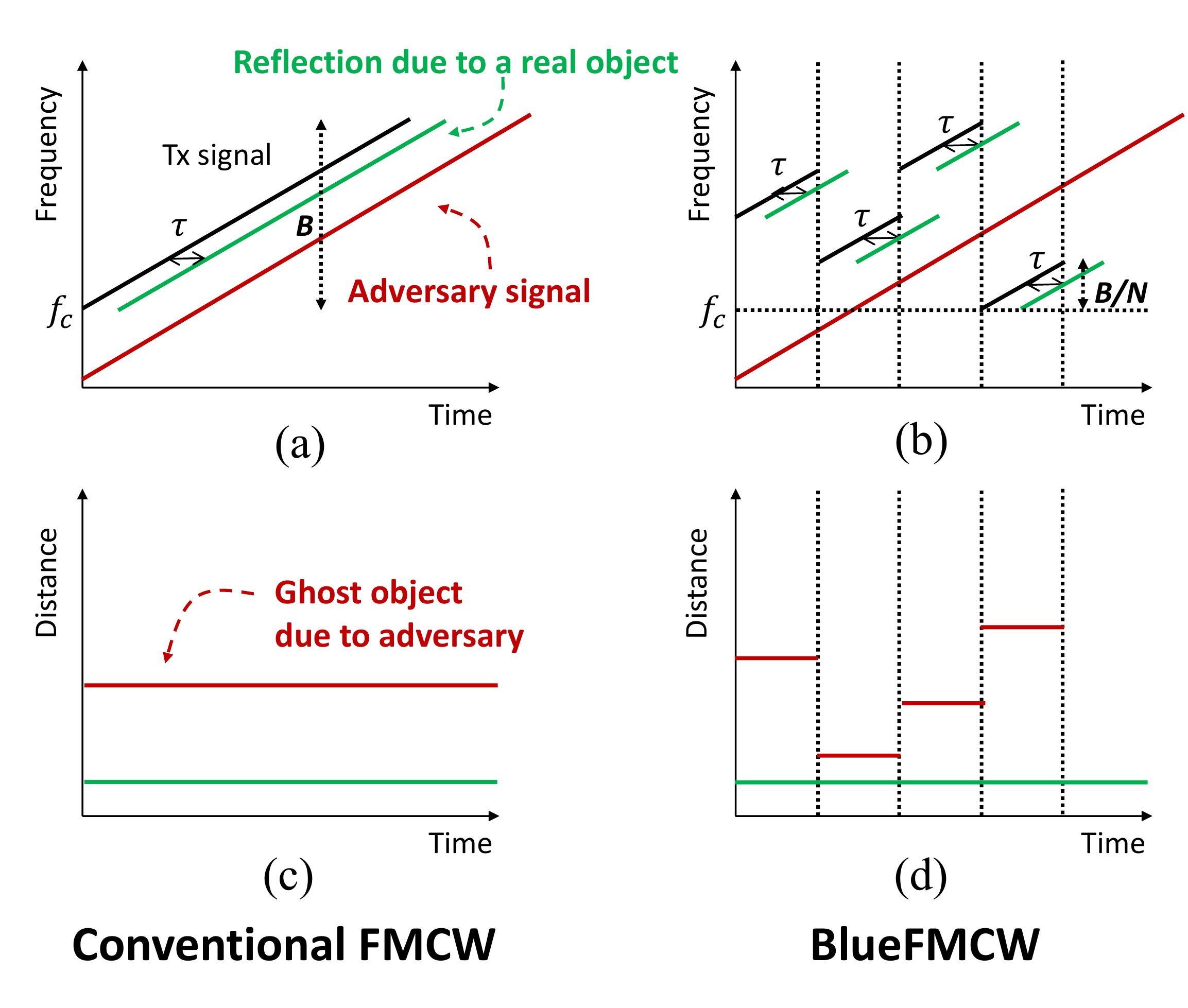}\\
\caption{ \textbf{Transmitted signal and beat signal of  (a)(c) conventional FMCW versus (b)(d) BlueFMCW .} } 
\label{fig:chirps}
\end{center}
\vskip -0.25in
\end{figure}

\subsection{Adversary models }
In this paper, we assume two adversary models. In the first model, a victim radar can be attacked by a spoofing signal intentionally generated by a malicious radar. The attacker can listen to the victim radar and replicate the identical signal to manipulate the ranging capability. Given that the technical standards of the conventional FMCW system (modulation type, carrier frequency, bandwidth, and pulse period) are known, the attacker can synchronize with the victim radar. \cite{miura2019low} recently demonstrates that the distance-spoofing attack on FMCW is feasible with low-cost off-the-shelf hardware. In the second model, two or more radars with identical parameters interfere with each other. It is assumed that a given model of vehicles would equip the same radar for the entire production. In this scenario, the interference is unintentional but can be accidentally synchronized with the other radars. 

In both scenarios, the victim radar would falsely detect an object, i.e. a ghost object. Figure~\ref{fig:chirps} (a) and (c) shows an example of the ghost object in the conventional FMCW radar. As the adversary signal (either by spoofing or interference) has the same slope of the original chirp signal, a false tone is created in the beat signal shown in  Figure~\ref{fig:chirps} (c). 

In this paper,  we will use the notation \textit{adversary signals} to represent both spoofing and interference signals that create a ghost object at the victim radar.

\section{BlueFMCW}
To address the above adversary scenarios, we introduce BlueFMCW, a radar system that mitigates the adversary signals. Specifically, instead of transmitting a full chirping signal from the lower to upper frequency, BlueFMCW makes a \textit{random-frequency-jump} in the middle of the chirp signal as shown in Figure~\ref{fig:chirps} (b).  While the TOF of the reflected signals remains the same, the frequency gap with the adversary signal will be randomized. As a result, the false beat frequency does not stay at the same FFT bin shown in Figure~\ref{fig:chirps} (d). In other words, the energy of the adversary signal will be \textit{randomly} dispersed over various FFT bins, which results in significantly smaller peaks in the spectrum. 

In this section, we first describe how to perform random frequency hopping in FMCW. We then address the challenges to reconstruct the beat signal. Finally, we discuss how BlueFMCW performs with different parameters and choose an optimal configuration. 

\subsection{Random Frequency Hopping}
\begin{figure*}[t]
\begin{center}
\includegraphics[width=\textwidth]{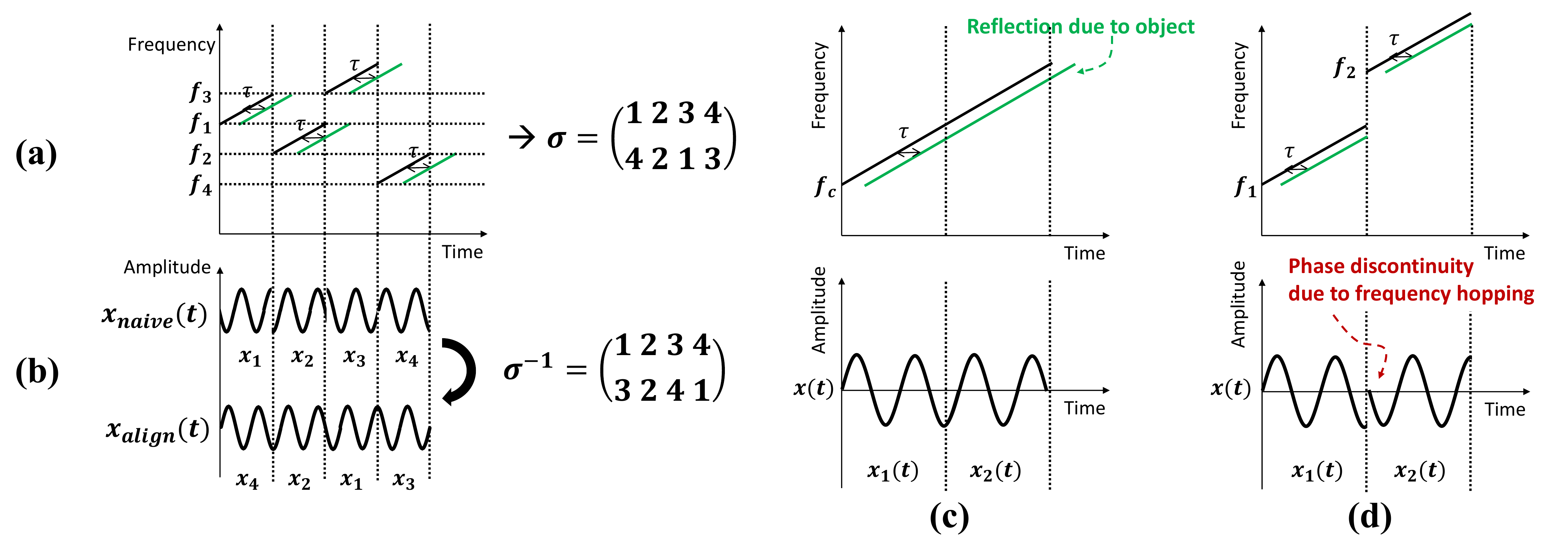}\\
\caption{ \textbf{Frequency hopping by permutation and its beat signal.} (a) shows an example of the transmitted BlueFMCW signal ($N=4$). The sub-chirps are permuted by $\sigma$. (b) shows the beat signals in the time-domain. $x_{align}$ is achieved by inverse-permuting $x_{naive}$. (c) shows the conventional FMCW and its beat signal. (d) shows an example of frequency hopping in BlueFMCW and the phase discontinuity in its beat signal. } 
\label{fig:perm}
\end{center}
\vskip -0.25in
\end{figure*}

Consider a conventional FMCW chirp signal, i.e., linear frequency modulation signal from a starting frequency $f_c$ for $T$ duration. BlueFMCW creates a series of frequency-hopping chirps by (1) dividing the conventional FMCW signal into $N$ equal sub-intervals, and then (2) randomly permuting the sub-chirps. Suppose the victim radar receives a reflected signal from a real object and an adversary signal that spawns a ghost target. Figure~\ref{fig:chirps} (c) and (d) shows the victim's spectrograms of the conventional FMCW and BlueFMCW. The beat frequency from the adversary signal remains constant through $T$ in the conventional FMCW. In contrast, BlueFMCW can \textit{hash} the beat frequency thanks to its \textit{randomized} staring frequencies of the sub-chirps. However, the beat frequency from the real object is not hashed and remains constant through $T$. This is because the beat frequency of a true object only depends on the TOF regardless of the starting frequency. BlueFMCW leverages this to mitigate the adversary signal while achieving the same detection capability on the real objects. 

To formalize this, we first denote the starting frequency of the $k$th sub-chirp to be $f_k$. For the random frequency hopping, we can define a random permutation $\sigma: F \rightarrow F$, where $F$ is a finite set of the index ${1, 2, ..., N}$. The two-line notation of $\sigma$ can be written as:
\begin{align}
\begin{pmatrix}
1 & 2 & ... & N\\
\sigma(1) & \sigma(2) & ... & \sigma(N)
\end{pmatrix}
\end{align}

Figure~\ref{fig:perm} (a) illustrates an example of BlueFMCW sub-chirps with $N=4$. In this example, the first sub-chirp of the conventional FMCW is permuted to the third slot, the second sub-chirp to the second slot, the third sub-chirp to the first slot, and the fourth sub-chirp to the third slot. In the next section, we will discuss the impact of the phase discontinuity of the beat signal and how we can reconstruct the original beat signal using the inverse permutation. 

\subsection{Reconstruction}
\subsubsection{Challenges: Resolution and Discontinuities}
By observing the pattern of the BlueFMCW spectrogram, one can speculate which one is the true object or adversary signal. The true object tends to have a consistent peak frequency while the adversary signal jumps from one to another frequency. Using this observation, one can estimate a rough distance profile from the single sub-chirp's beat signal. However, the distance resolution of a sub-chirp is degraded by $N$ because its bandwidth is reduced by $B/N$. As discussed in Section~\ref{sec_background}, the distance resolution achieved by a single sub-chirp becomes $\frac{cN}{2B}$. 

A naive attempt to solve the resolution problem is simply concatenating the beat signals of all sub-chirps in time-order such as
\begin{align}
x_{naive}(t) = [x_1(t), x_2(t), ..., x_N(t)]
\end{align}
where $x_k(t)$ is the beat signal by $k$-th sub-chirp. By concatenating all $N$ beat signals as illustrated in Figure~\ref{fig:perm} (b), the resolution will remain the same with the conventional FMCW, $\frac{c}{2B}$. It will create, however, spurious frequency components in the frequency domain. To understand why, we need to examine the phase of the beat signal. Recall the beat signal from Eq.~\ref{eq:beat}. For simplicity, we can rewrite the beat signal of the $k$-th sub-chirp with a single reflection:
\begin{align} \label{eq:sub}
x_k(t) = \exp(j2\pi (\alpha \tau t + f_k \tau - \alpha \tau^2 /2)) \quad ,t \in [0, T/N]
\end{align}
The phase in frequency is written as:
\begin{align}
\phi =  2\pi(f_k \tau - \alpha \tau^2 /2)
\end{align}

Figure~\ref{fig:perm} (c) shows an example of the conventional FMCW.  As can be seen, the conventional FMCW signal is a special case of BlueFMCW where $f_k$'s are sorted in ascending order. Hence, we can represent $x_2(t)$ in two ways; one with Eq.~\ref{eq:sub} and the other with the $T/N$ advanced version of Eq.~\ref{eq:beat}. 
\begin{align}
x_2(t) = 
\begin{cases}
\exp(j2\pi (\alpha \tau t + f_2 \tau - \alpha \tau^2 /2)) & \text{by Eq.~\ref{eq:sub}}\\
\exp(j2\pi (\alpha \tau (t+T/N) + f_1 \tau - \alpha \tau^2 /2)) & \text{by Eq.~\ref{eq:beat}}
\end{cases}
\end{align}
For the conventional FMCW, the phases of the two equation always the same.
\begin{align}
2\pi(f_2 \tau - \alpha \tau^2 /2) = 2\pi(\alpha \tau T/N + f_1 \tau - \alpha \tau^2 /2)
\end{align}
The two phases are always identical in the conventional FMCW because $f_2$ is equal to $f_1 + \alpha T/N$. 

However, if the sub-chirps are randomly permuted, it is not guaranteed that $f_{k+1}$ is equal to $f_k + \alpha T/N$. This will cause phase discontinuities as shown in Figure~\ref{fig:perm} (d). We will address this issue in the following section. 

\subsubsection{Phase Alignment} \label{sec_PA}
\begin{figure}[t]
\begin{center}
\includegraphics[width=0.43\textwidth]{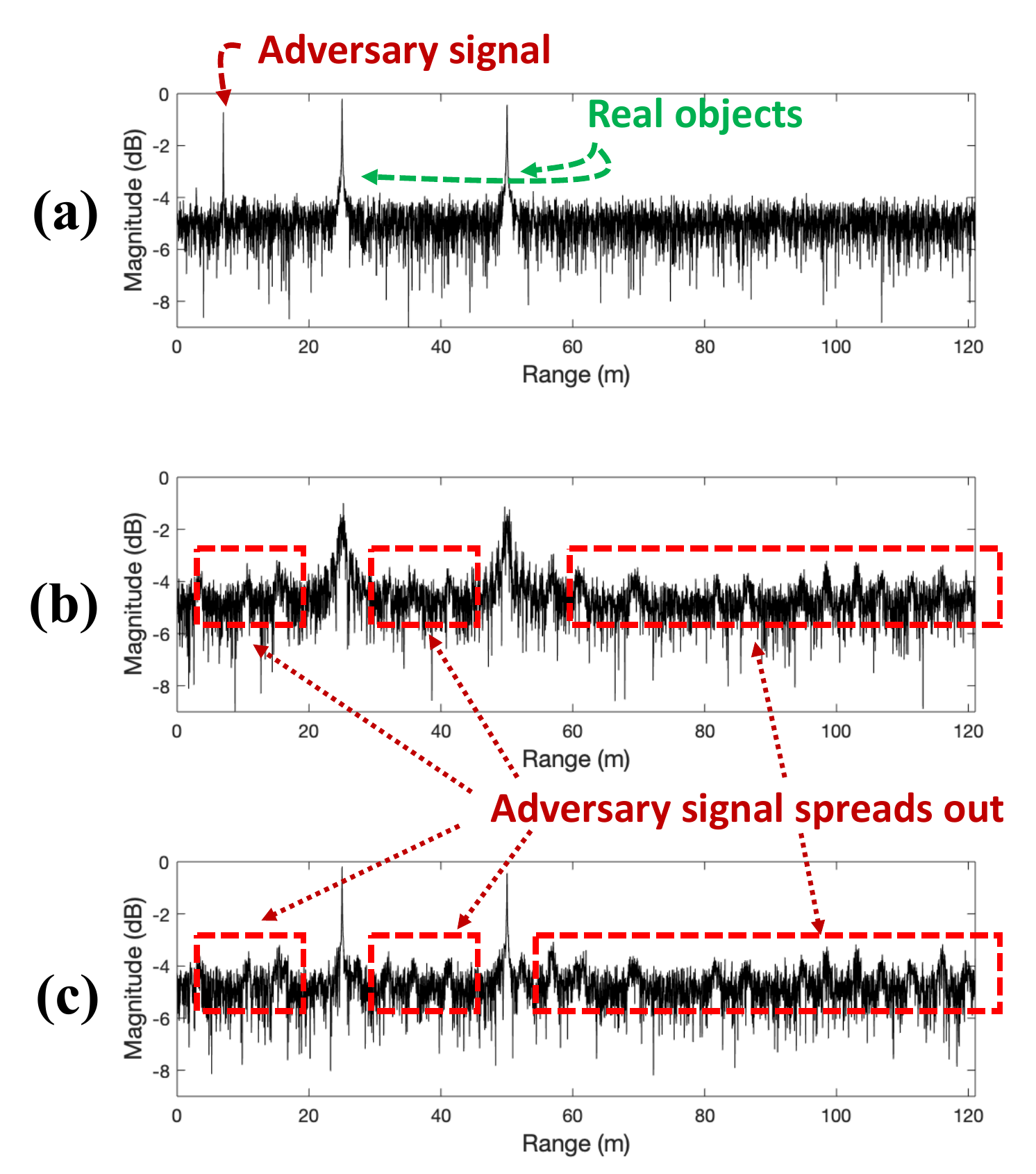}\\
\caption{ \textbf{Beat spectrum (distance profile) of (a) conventional FMCW, (b) BlueFMCW wihtout phase alignment, and (c) BlueFMCW with phase alignment.} } 
\label{fig:cmp}
\end{center}
\vskip -0.25in
\end{figure}
To eliminate the phase discontinuity due to the frequency hopping, BlueFMCW re-arranges the beat signals by inverting the permutation used in the frequency hopping. For example in Figure~\ref{fig:perm} (a), the first sub-chirp of the conventional FMCW (starting frequency $f_1$) was permuted to the fourth time-slot. If we concatenate the beat signals in time-order, the beat signal starting at $f_1$ comes at the last place, causing the phase discontinuity. As we know how the sub-chirp was permuted ($\sigma$), we can invert the permutation to bring the beat signal back to the correct time-slot so that the phase becomes continuous. By the inverse permutation $\sigma^{-1}$, the fourth beat signal is permuted to the first time-slot, and so the others (Figure~\ref{fig:perm} (b)). 

Formally, the phase aligned beat signal, $x_{align}$, is achieved by performing the inverse permutation $\sigma^{-1}$ on the time-order beat signals, $x_{naive}$. 
\begin{align}
x_{naive}(t) &= [x_1, x_2, ..., x_N ]\\
x_{align}(t) &= [x_{\sigma^{-1}(1)}, x_{\sigma^{-1}(2)}, ..., x_{\sigma^{-1}(N)}]
\end{align}

Figure~\ref{fig:cmp} shows the impact of the phase alignment in the beat spectrum. In this example, two true objects and one adversary signal are simulated. The conventional FMCW in Figure~\ref{fig:cmp} (a) images all three components in the spectrum. In Figure~\ref{fig:cmp} (b), the spectrum of $x_{naive}$ is shown. Due to the phase discontinuity, we can observe many spurious signals around the objects and the smaller and unclear peaks for the true objects. After the phase alignment, Figure~\ref{fig:cmp} (c) shows two clear peaks of the objects. Note that both BlueFMCWs with and without phase alignment can alleviate the adversary signal. This is because random frequency hopping makes random frequency gaps against the adversary, and thus the result beat frequencies also become randomized whether phase-aligned or not. The mitigated adversary signal, however, did not disappear. Instead of persisting at one frequency, the adversary signal is randomly dispersed over the various frequencies. 

\subsection{Designing BlueFMCW} \label{sec_design}
\begin{figure}[t]
\begin{center}
\includegraphics[width=0.4\textwidth]{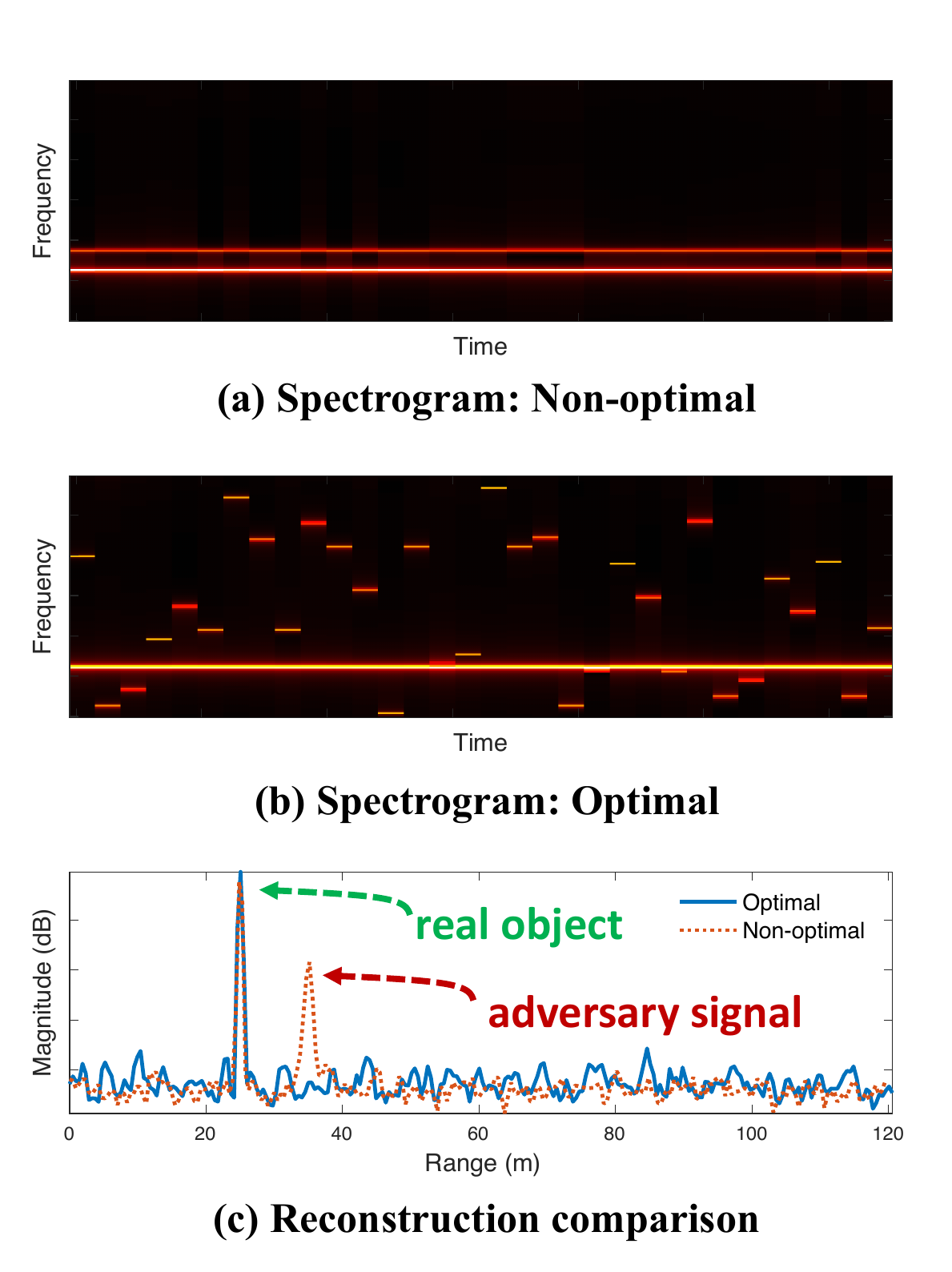}\\
\caption{ \textbf{Comparing the spectrogram of the beat signal with (a) non-optimal configuration and (b) optimal configuration. The beat spectrum results are compared in (c).}  } 
\label{fig:para}
\end{center}
\vskip -0.25in
\end{figure}
In the previous section, we showed that BlueFMCW can spread out the adversary signal while holding the phase continuity of the reflected signals from the real objects. We assume the sub-chirps share the same slope ($\alpha$) and duration ($T_b$) for the simplicity. In addition, the frequency support of each sub-chirp must be mutual exclusive in order to align the phase by the permutation. For example, if a frequency support overlaps with another, the phase cannot be aligned without dropping the overlapping samples. For the opposite case, if there are empty gaps among the frequency supports, the phase alignment again cannot be achieved without interpolating the missing samples. Thus, having the frequency supports to be mutual exclusive each other and to cover the full bandwidth makes the problems easy. 

The question remains \textit{how diverse BlueFMCW can spread out the adversary signals}. To maximize the spread, we want to avoid the adversary signals falling in the same FFT bin. By properly choosing the sub-chirp parameters, the down-converted adversary signals fall into many different FFT bins.\footnote{In order to highlight the impact of sub-chirp design, we assume a worst-case scenario that the out-of-band adversary signal after the receiver mixer can be leaked to ADC and cause aliasing spectrum in the FFT. In practice, the LPF can filter out most of the out-of-band signals.} As a result, the energy of the adversary signal will be widely spread over the bandwidth. In the opposite case, when the adversary signals fall at a few identical FFT bins, they will end up with high-magnitude tones in FFT. 

The beat frequency of an adversary signal in FFT can be represented by the starting frequencies of the victim and aggressor (interferer or attacker), the bandwidth of the sub-chirps ($B_{sub} = \alpha T_b$), and the sampling rate ($f_s$). 
\begin{align} \label{eq:fbeat}
f_{beat} &= mod(\Delta f  + k B_{sub}, f_s) \quad , k = 0,1,2...,N-1
\end{align}
where $\Delta f$ is the frequency difference between the transmitted signal and the received adversary signal, and $k$ is an integer dependent on the permutation. In the conventional FMCW, $m$ is zero for the entire chirp duration which makes the beat frequency constant. Since $\Delta f$ is dependent on the distance, the starting frequency of the aggressor, and the delay of the adversary signal, BlueFMCW does not have control over it. However, BlueFMCW can configure $B_{sub}$ and $f_s$. Assume they are set to $B_{sub} = \frac{n}{m} f_s$ where $m$ and $n$ are the integers. To calculate $f_{beat}$, consider the argument of the modulo operation in Eq.~\ref{eq:fbeat} as
\begin{align}
\frac{\Delta f  + k B_{sub}}{f_s} & = \frac{\Delta f}{f_s} + \frac{kn}{m} \\
                                    & = \underbrace{Q}_{\text{Quotient}} + \underbrace{\delta f + \frac{p}{m}}_{\text{remainder}} 
\end{align}
where $Q$ is the quotient of the division, and $\delta f + \frac{p}{m}$ is the remainder where $p$ is an integer such that $-m<p<m$.  We can use the above result to rewrite the beat frequency as
\begin{align}
f_{beat} &= (\delta f + \frac{p}{m}) f_s
\end{align}
Note that $\delta f$ is the constant remainder from $\frac{\Delta f}{f_s}$ that does not impact on the diversity. Therefore, there exists at most $m$ different $f_{beat}$. For the worst case example ($m=1$), consider $N=4$, $f_s=100MHz$, $B_{sub} = 2 f_s = 200MHz$, and $\Delta f = 70MHz$. No matter how we hop the frequency, the remainder is $70/100$, and thereby $f_{beat}$ is always 70MHz. For a better configuration ($m=3$), consider $B_{sub} = \frac{7}{3} f_s$ with the same setup for the rest. For $k=0,1,2,3$, the remainders are $70/100$, $70/100 + 1/3$, $70/100 + 2/3$, and $70/100$, respectively. Therefore, we can achieve 3 different $f_{beat}$ as $m$ is set to 3. 

Figure~\ref{fig:para} (a) shows the BlueFMCW spectrogram of the worst case. Clearly, the adversary is not spread even with the random frequency hopping. This is because $f_s$ and $B_{sub}$ have the integer relationship with $m=1$. The red lines in Figure~\ref{fig:para} (c) correspond to the reconstruction result by the worst case. Figure~\ref{fig:para} (b) shows the BlueFMCW spectrogram of $m=131$. Compared to the worst case, the adversary signal is spread on 131 different beat frequencies. As a result, the adversary signal is greatly reduced as shown in the blue lines in Figure~\ref{fig:para} (c). 

\section{Results}
\begin{table}
\renewcommand{\arraystretch}{1.3}
\caption{Experimental Setup}
\label{tab:config}
\centering
\begin{tabular}{ c c c } 
\hline
 & Parameter & Value\\ 
\hline
\multirow{6}{*}{FMCW}   & $f_c$ & 24 GHz \\
                            & Sampling rate & 20 MHz \\
                            & Chirp duration & 204.8 $\mu$s\\
                            & \# of total samples & 4096\\
                            & Slope (optimal) & 24.785 MHz/$\mu$s\\
                            & Slope (non-optimal) & 26.562 MHz/$\mu$s\\                            
\hline
\multirow{2}{*}{BlueFMCW}  & \# of sub-chirps & 32 \\
                            & Sub-chirp duration & 6.4 $\mu$s\\
\hline
\multirow{3}{*}{Aggressor}  & \# of Aggressors & 1-10 \\
                            & Distance (m) & Unif(0.5, 200)\\
                            & SIR(dB) & Unif(2.5, 6)\\
\hline
Static object   & Distance & 25 m\\
\hline
\end{tabular}
\end{table}

\subsection{Experimental Setup}
We conducted a simulation on Matlab to evaluate BlueFMCW's ability to mitigate the adversary signals. The victim radar is configured with the starting frequency 24GHz, 20MHz sampling rate, and 4096 number of samples. We choose two slope rates of the chirp; 24.785 MHz/us for the optimal configuration and 26.562 MHz/us for the non-optimal. In this simulation, BlueFMCW divides the chirp into 32 sub-chirps and randomly permutes them while sharing the common FMCW parameters. The static object is located at distance of 25 meters. We simulated 10 different numbers of the aggressor radars, 1-10 aggressors. The distance to the aggressors is randomly chosen between 0.5-200 meters. The power of the aggressor signal is also randomly generated to have SIR between 2.5-6 dB. Table~\ref{tab:config} summarizes the experimental setup. 

\subsection{Evaluation Metrics}
We evaluate the performance of BlueFMCW along two axes. The first is how well it can mitigate the adversary. In this case, our metric is the \textit{signal-to-interference ratio (SIR)} defined by the ratio between the signal power of the true object and the strongest interference signal seen in the beat frequency domain. The higher the SIR, the better BlueFMCW mitigates the adversary signals. 

The second metric is the \textit{signal-to-interference-plus-noise-ratio (SINR)} loss. BlueFMCW spreads the adversary signals across the bandwidth. This will end up increasing noise floor, i.e. decreasing SINR. We first calculate the SINR of the conventional FMCW without any adversary signals, $SINR_{base}$. Then, we compare $SINR_{base}$ with the SINR degraded by adversary signals. 
\begin{align}
SINR_{loss\_conv} = SINR_{base} - SINR_{conv}\\
SINR_{loss\_blue} = SINR_{base} - SINR_{blue}
\end{align}

\subsection{Compare Scenarios}
We compare the following scenarios:
\begin{itemize}
    \item \textbf{Conventional FMCW vs Conventioanl FMCW's (CvC): } The victim radar and aggressor radars (interferers or attackers) use the conventional FMCW. 
    \item \textbf{BlueFMCW vs Conventioanl FMCW's (BvC): } The victim radar uses BlueFMCW, but the aggressor radars use the conventional FMCW. 
    \item \textbf{BlueFMCW vs BlueFMCW's (BvB): } The victim radar and aggressor radars use BlueFMCW. However, they do not share the random permutation. 
\end{itemize}

In addition, we evaluate the impact of the phase alignment and the BlueFMCW design on the performance. 

\subsection{Adversary Mitigation Performance} 
\begin{figure}[t]
\begin{center}
\includegraphics[width=0.4\textwidth]{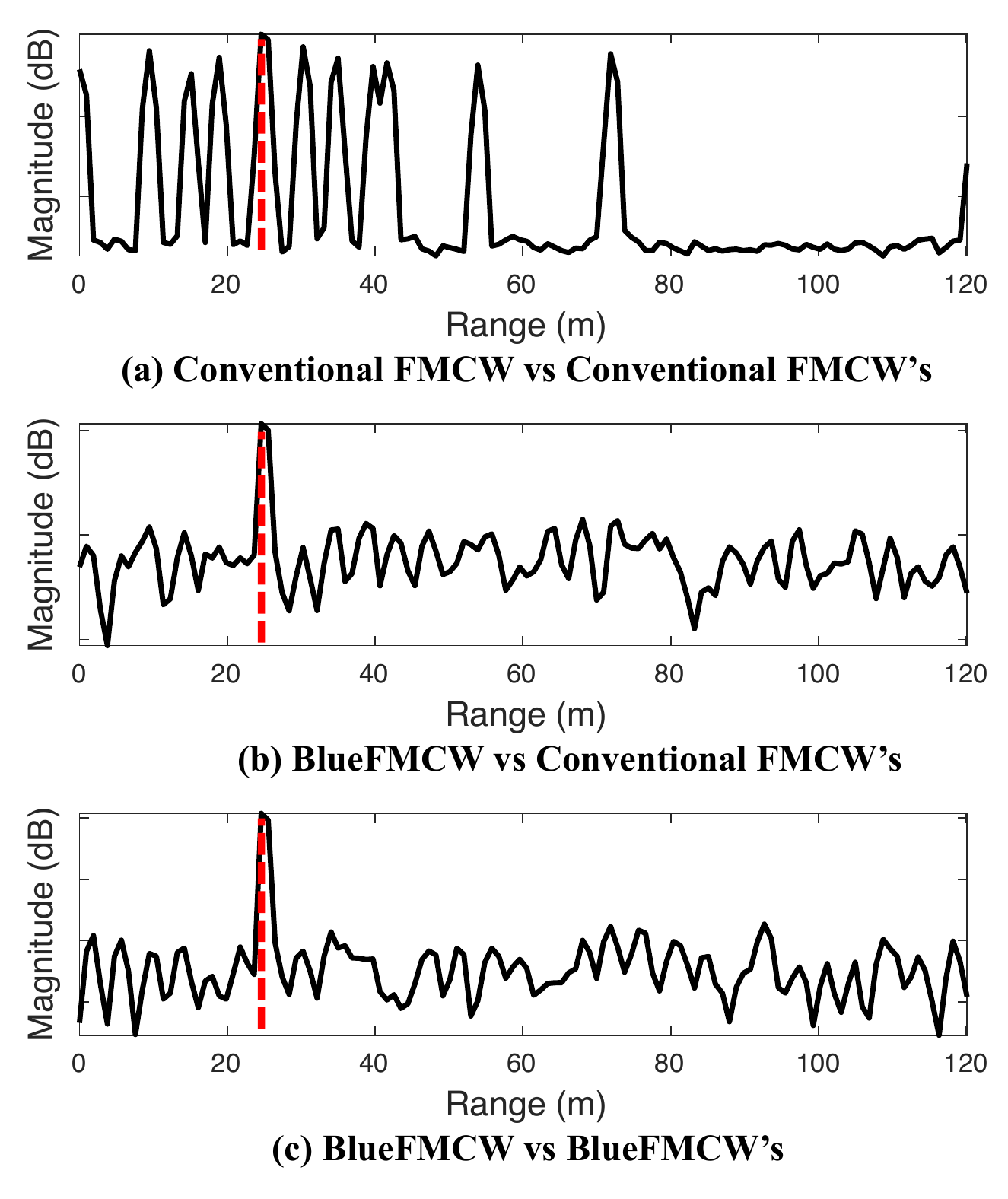}\\
\caption{ \textbf{Beat spectrum under (a) CvC (b) BvC (c) BvB scenario.}} 
\label{fig:allspect}
\end{center}
\vskip -0.25in
\end{figure}

\begin{figure}[t]
\begin{center}
\includegraphics[width=0.4\textwidth]{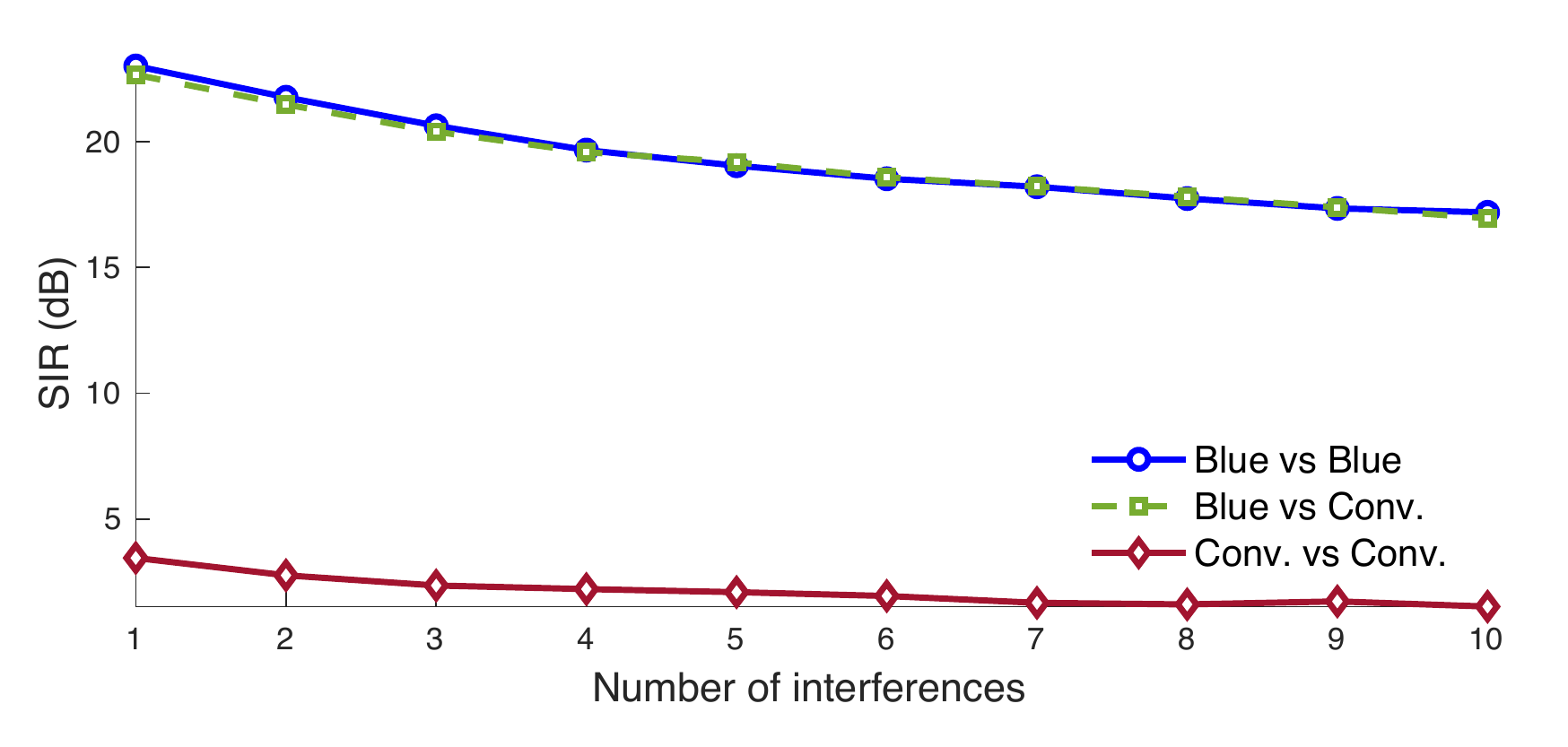}\\
\caption{ \textbf{ 50$^{th}$ percentile SIR as the number of adversary signals varies.} } 
\label{fig:sirvn}
\end{center}
\vskip -0.25in
\end{figure}

\begin{figure}[t]
\begin{center}
\includegraphics[width=0.4\textwidth]{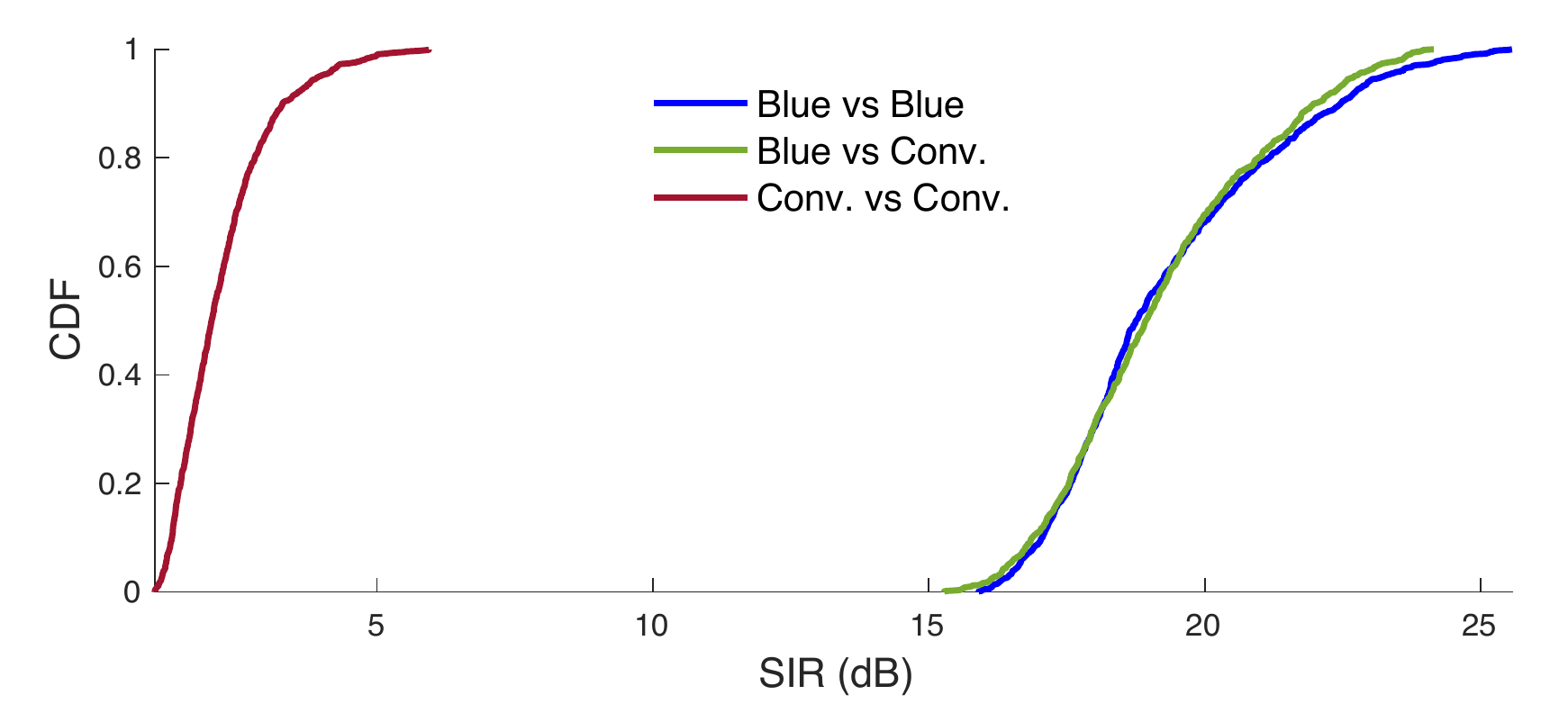}\\
\caption{ \textbf{SIR for different scenarios.} } 
\label{fig:sircdf}
\end{center}
\vskip -0.25in
\end{figure}

\noindent \textbf{Impact of the number of attackers: } We first demonstrate the impact of the number of adversary signals on the  performance of BlueFMCW. We compare the above three scenarios across the number of adversaries. The distance of the aggressor radars and the power of the adversary signals follow the uniform distribution in Table~\ref{tab:config}. Each case repeats 100 runs. One of the beat spectrum of the three scenarios with 10 adversary signals are compared in Fig~\ref{fig:allspect}. The red dotted line represents the distance of the real object. As shown in Fig~\ref{fig:allspect} (a), the victim radar with the first scenario (all radars are conventional FMCW) has no capability of mitigating the adversary signals. However, when the victim radar uses BlueFMCW, it can suppress the adversary significantly regardless of the aggressor radar type as shown in Fig~\ref{fig:allspect} (b) and (c). Fig~\ref{fig:sirvn} shows the 50$^{th}$ percentile SIR as a function of the number of adversary signals. As the number of adversaries increases, BlueFMCW is able to uphold the high SIR in both BvC and BvB scenarios. On the otherhand, the conventional FMCW is unable to mitigate a single adversary signal. Fig~\ref{fig:sircdf} plots the CDF of the SIRs of the three scenarios. BlueFMCW can achieve the 50$^{th}$ percentile SIR of 18.75 dB for BvB scenario and 18.93dB for BvC scenario. Whereas in CvC scenario, the 50$^{th}$ percentile SIR is around 2dB.

\begin{figure}[t]
\begin{center}
\includegraphics[width=0.4\textwidth]{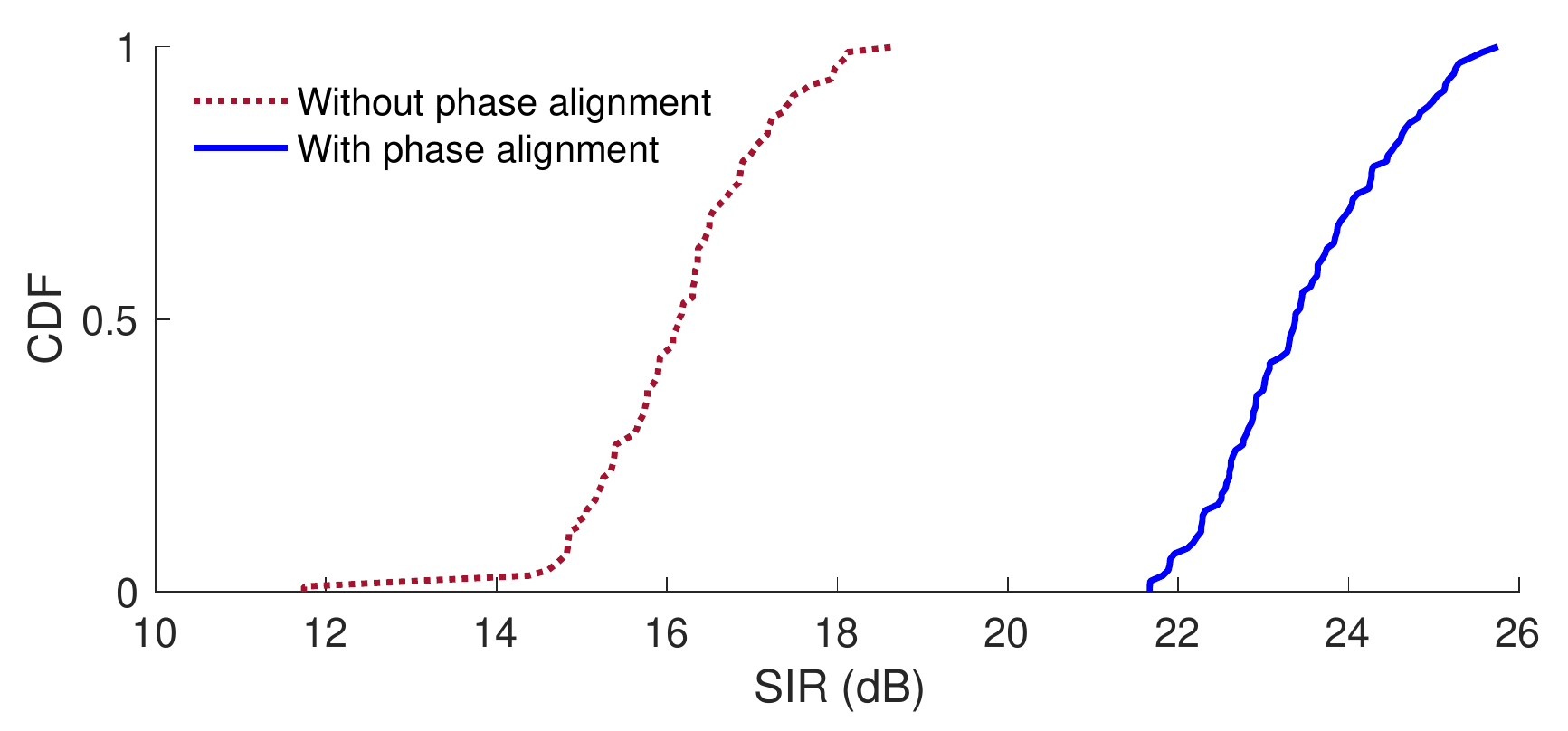}\\
\caption{ \textbf{SIR with and without phase alignment.} } 
\label{fig:sirpa}
\end{center}
\vskip -0.25in
\end{figure}

\begin{figure}[t]
\begin{center}
\includegraphics[width=0.4\textwidth]{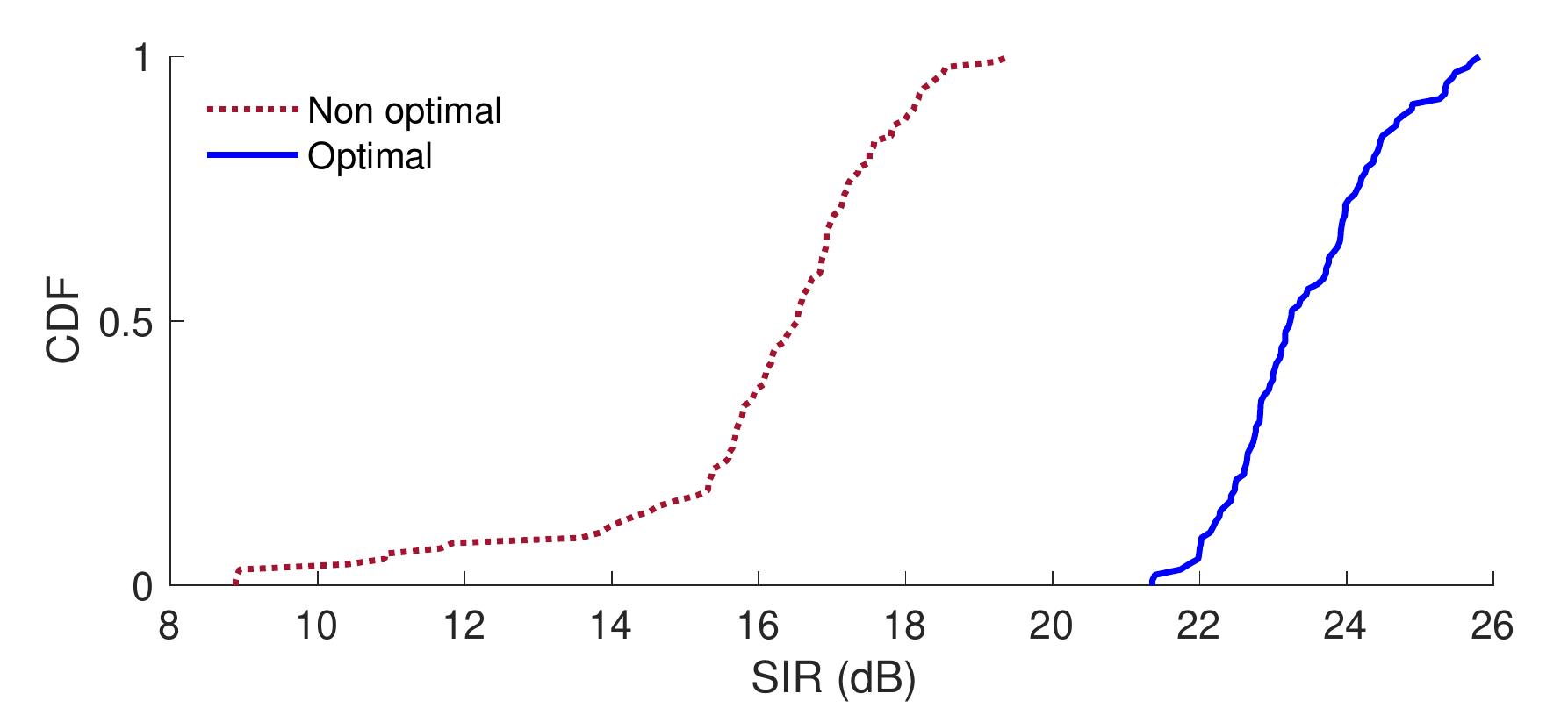}\\
\caption{ \textbf{SIR with optimal and non-optimal configuration.}} 
\label{fig:siropt}
\end{center}
\vskip -0.25in
\end{figure}

\begin{figure}[t]
\begin{center}
\includegraphics[width=0.4\textwidth]{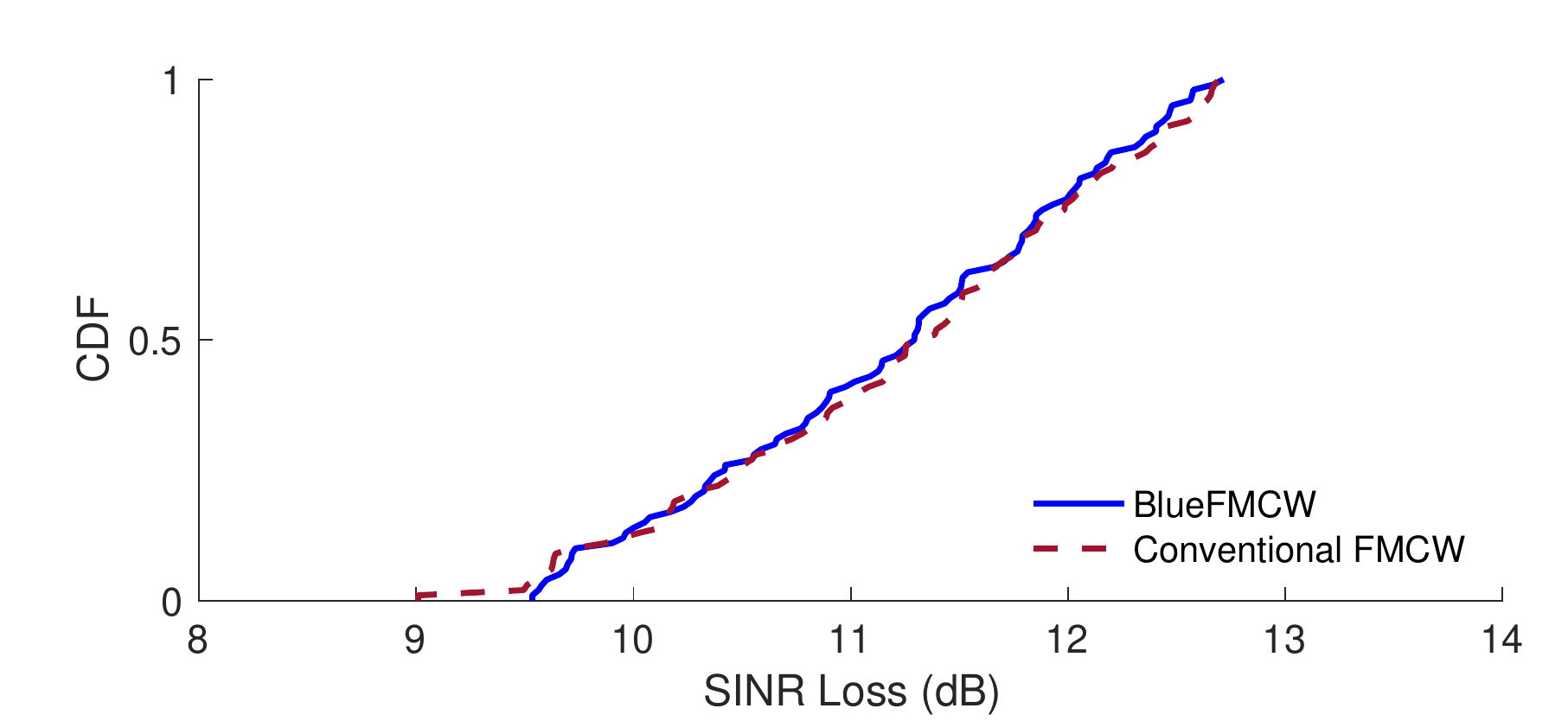}\\
\caption{ \textbf{SINR loss of BlueFMCW vs conventional FMCW.}} 
\label{fig:sinr}
\end{center}
\vskip -0.25in
\end{figure}

\noindent \textbf{Impact of the phase alignment: } Next we evaluate the impact of the phase alignment on the performance of BlueFMCW. In Section~\ref{sec_PA}, we explain that the phase alignment reduces the discontinuities as combining the received signals to improve the distance resolution. Fig~\ref{fig:sirpa} plots the CDF of SIR gain with and without the phase alignment. The figure shows that the phase alignment can improve SIR more than 7dB. This is because the phase discontinuity is minimized by the alignment, and therefore there are less spurs introduced by the true object signals. 

\noindent \textbf{Optimal vs Non-optimal configuration: } In this experiment, we evaluate the performance of BlueFMCW with different chirp parameters. As explained in Section~\ref{sec_design}, we can configure BlueFMCW chirp to spread the adversary signal on more or less distinctive beat frequencies, i.e. how well it can disperse the adversary on the noise floor. Spreading the adversary signal on more number of distinctive beat frequencies will increase SIR. Fig~\ref{fig:siropt} compares the CDF of SIR when BlueFMCW spreads the adversary signal on (i) 131 possible beat frequencies (optimal configuration), and (ii) 2 possible beat frequencies (non-optimal configuration). We only changed the slope of the chirp while fixing the rest configuration as shown in Table~\ref{tab:config}. This will effectively change the bandwidth of the FMCW chirp. The non-optimal configuration degrades the median SIR by 7 dB and the 10$^{th}$ percentile SIR by 9 dB. This is due to the fact that there are only two possible beat frequencies where the adversary energy falls onto. In contrast, the optimal configuration can create 131 possible beat frequencies and the adversary signals have more location to be spread out. 

\subsection{SINR Degradation} So far, we focus on the adversary mitigation performance of BlueFMCW by measuring SIR gain. In this experiment, we want to understand how the dispersed adversary signal will impact on the noise level. We first calculate the SINR of the conventional FMCW without any adversary, which serves as the base line. Then, we compare the base SINR with the SINR by the conventional FMCW and BlueFMCW interfered by an aggressor. Fig~\ref{fig:sinr} compares the CDF of SINR loss of the conventional FMCW and BlueFMCW. The figure shows both cases have almost same amount of the SINR loss. This suggests that the energy of the adversary is not either created nor destroyed. BlueFMCW spreads the energy of the adversary into many frequencies so that none of them creates a strong peak. This will effectively increase the noise floor which is to pay in BlueFMCW. However, increasing the noise floor is a less critical problem than detecting a ghost object. 
\section{Conclusion}
In this paper, we introduced BlueFMCW, a novel radar system that can efficiently mitigate both spoofing and interference signals without compromising radar's resolution. Our simulation results demonstrated that BlueFMCW can significantly improve the SINR ratio compared to the conventional FMCW system. Moving forward, we are interested in implementing BlueFMCW in hardware to prove its feasibility. 

\bibliographystyle{IEEEtran}
\bibliography{access}

\begin{thebibliography}{10}
\providecommand{\url}[1]{#1}
\csname url@samestyle\endcsname
\providecommand{\newblock}{\relax}
\providecommand{\bibinfo}[2]{#2}
\providecommand{\BIBentrySTDinterwordspacing}{\spaceskip=0pt\relax}
\providecommand{\BIBentryALTinterwordstretchfactor}{4}
\providecommand{\BIBentryALTinterwordspacing}{\spaceskip=\fontdimen2\font plus
\BIBentryALTinterwordstretchfactor\fontdimen3\font minus
  \fontdimen4\font\relax}
\providecommand{\BIBforeignlanguage}[2]{{%
\expandafter\ifx\csname l@#1\endcsname\relax
\typeout{** WARNING: IEEEtran.bst: No hyphenation pattern has been}%
\typeout{** loaded for the language `#1'. Using the pattern for}%
\typeout{** the default language instead.}%
\else
\language=\csname l@#1\endcsname
\fi
#2}}
\providecommand{\BIBdecl}{\relax}
\BIBdecl

\bibitem{radaridentity}
J.~{Dai}, X.~{Hao}, P.~{Li}, Z.~{Li}, and X.~{Yan}, ``Antijamming design and
  analysis of a novel pulse compression radar signal based on radar identity
  and chaotic encryption,'' \emph{IEEE Access}, vol.~8, pp. 5873--5884, 2020.

\bibitem{FMCWLidar}
\BIBentryALTinterwordspacing
F.~Zhang, L.~Yi, and X.~Qu, ``Simultaneous measurements of velocity and
  distance via a dual-path fmcw lidar system,'' \emph{Optics Communications},
  vol. 474, p. 126066, 2020. [Online]. Available:
  \url{http://www.sciencedirect.com/science/article/pii/S0030401820304831}
\BIBentrySTDinterwordspacing

\bibitem{FMCWLidarInterference}
\BIBentryALTinterwordspacing
I.-P. Hwang, S.-J. Yun, and C.-H. Lee, ``Mutual interferences in
  frequency-modulated continuous-wave (fmcw) lidars,'' \emph{Optik}, p. 165109,
  2020. [Online]. Available:
  \url{http://www.sciencedirect.com/science/article/pii/S0030402620309451}
\BIBentrySTDinterwordspacing

\bibitem{alland2019interference}
S.~Alland, W.~Stark, M.~Ali, and M.~Hegde, ``Interference in automotive radar
  systems: Characteristics, mitigation techniques, and current and future
  research,'' \emph{IEEE Signal Processing Magazine}, vol.~36, no.~5, pp.
  45--59, 2019.

\bibitem{litman2017autonomous}
T.~Litman, \emph{Autonomous vehicle implementation predictions}.\hskip 1em plus
  0.5em minus 0.4em\relax Victoria Transport Policy Institute Victoria, Canada,
  2017.

\bibitem{heath}
E.~Yeh, J.~Choi, N.~Prelcic, C.~Bhat, and R.~W. Heath~Jr, ``Security in
  automotive radar and vehicular networks,'' \emph{submitted to Microwave
  Journal}, 2016.

\bibitem{chaun14}
R.~{Chauhan}, ``A platform for false data injection in frequency modulated
  continuous wave radar,'' [Online]. Available:
  \url{http://digitalcommons.usu.edu/cgi/viewcontent.cgi?article=4983&context=etd}.

\bibitem{ma2019joint}
D.~Ma, N.~Shlezinger, T.~Huang, Y.~Liu, and Y.~C. Eldar, ``Joint
  radar-communications strategies for autonomous vehicles,'' \emph{arXiv
  preprint arXiv:1909.01729}, 2019.

\bibitem{huang2020majorcom}
T.~Huang, N.~Shlezinger, X.~Xu, Y.~Liu, and Y.~C. Eldar, ``Majorcom: A
  dual-function radar communication system using index modulation,'' \emph{IEEE
  Transactions on Signal Processing}, 2020.

\bibitem{bechter2016digital}
J.~Bechter, K.~Eid, F.~Roos, and C.~Waldschmidt, ``Digital beamforming to
  mitigate automotive radar interference,'' in \emph{2016 IEEE MTT-S
  International Conference on Microwaves for Intelligent Mobility
  (ICMIM)}.\hskip 1em plus 0.5em minus 0.4em\relax IEEE, 2016, pp. 1--4.

\bibitem{bechter2017analytical}
J.~Bechter, M.~Rameez, and C.~Waldschmidt, ``Analytical and experimental
  investigations on mitigation of interference in a dbf mimo radar,''
  \emph{IEEE Transactions on Microwave Theory and Techniques}, vol.~65, no.~5,
  pp. 1727--1734, 2017.

\bibitem{dai2012main}
H.~Dai, X.~Wang, Y.~Li, Y.~Liu, and S.~Xiao, ``Main-lobe jamming suppression
  method of using spatial polarization characteristics of antennas,''
  \emph{IEEE Transactions on Aerospace and Electronic Systems}, vol.~48, no.~3,
  pp. 2167--2179, 2012.

\bibitem{dai2011novel}
H.~Dai, X.~Wang, and Y.~Li, ``Novel discrimination method of digital deceptive
  jamming in mono-pulse radar,'' \emph{Journal of Systems Engineering and
  Electronics}, vol.~22, no.~6, pp. 910--916, 2011.

\bibitem{gerstmair2019safe}
M.~Gerstmair, A.~Melzer, A.~Onic, and M.~Huemer, ``On the safe road toward
  autonomous driving: Phase noise monitoring in radar sensors for functional
  safety compliance,'' \emph{IEEE Signal Processing Magazine}, vol.~36, no.~5,
  pp. 60--70, 2019.

\bibitem{jin2019automotive}
F.~Jin and S.~Cao, ``Automotive radar interference mitigation using adaptive
  noise canceller,'' \emph{IEEE Transactions on Vehicular Technology}, vol.~68,
  no.~4, pp. 3747--3754, 2019.

\bibitem{uysal2019pcfmcw}
F.~{Uysal}, ``Phase-coded fmcw automotive radar: System design and interference
  mitigation,'' \emph{IEEE Transactions on Vehicular Technology}, vol.~69,
  no.~1, pp. 270--281, 2020.

\bibitem{bluetooth}
J.~C. {Haartsen}, ``The bluetooth radio system,'' \emph{IEEE Personal
  Communications}, vol.~7, no.~1, pp. 28--36, 2000.

\bibitem{guan2019application}
Z.~Guan, Y.~Chen, P.~Lei, D.~Li, and Y.~Zhao, ``Application of hash function on
  fmcw based millimeter-wave radar against drfm jamming,'' \emph{IEEE Access},
  vol.~7, pp. 92\,285--92\,295, 2019.

\bibitem{liu2018high}
J.~Liu, Y.~Zhang, and X.~Dong, ``High resolution moving train imaging using
  linear-fm random radar waveform,'' in \emph{2018 Asia-Pacific Microwave
  Conference (APMC)}.\hskip 1em plus 0.5em minus 0.4em\relax IEEE, 2018, pp.
  839--841.

\bibitem{miura2019low}
N.~Miura, T.~Machida, K.~Matsuda, M.~Nagata, S.~Nashimoto, and D.~Suzuki, ``A
  low-cost replica-based distance-spoofing attack on mmwave fmcw radar,'' in
  \emph{Proceedings of the 3rd ACM Workshop on Attacks and Solutions in
  Hardware Security Workshop}, 2019, pp. 95--100.

\end{thebibliography}



\end{document}